\documentclass[manuscript]{aastex}
\tighten
\newcommand\be{\begin{equation}}
\newcommand\ee{\end{equation}}

\usepackage{epstopdf}

\begin{document}
\shortauthors{Verma et al.}
\shorttitle{Acoustic Glitches in Main-Sequence Stellar models}

\title{A Theoretical Study of Acoustic Glitches in Low-Mass Main-Sequence Stars}
\author{Kuldeep Verma, H. M. Antia}
\affil{Tata Institute of Fundamental Research,
Homi Bhabha Road, Mumbai 400005, India}
\email{kuldeepv@tifr.res.in, antia@tifr.res.in}
\and
\author{Sarbani Basu}
\affil{Department of Astronomy, Yale University, P. O. Box 208101,
New Haven CT 06520-8101, U.S.A.}
\email{sarbani.basu@yale.edu}
\and
\author{Anwesh Mazumdar}
\affil{Homi Bhabha Centre for Science Education, Tata Institute of Fundamental Research,
V. N. Purav Marg, Mankhurd, Mumbai 400088, India}
\email{anwesh@tifr.res.in}

\begin{abstract}
There are regions in stars, such as ionization zones and the interface between
radiative and convective regions, that cause a localized sharp variation
in the sound speed. These are known as ``acoustic glitches''. Acoustic glitches
leave their signatures on the oscillation frequencies of stars, and hence
these signature can be used as diagnostics of these regions. In particular, the signature of these glitches
can be used as diagnostics of the position of the second helium ionization zone and that of the
base of the envelope convection zone. With the help of stellar models we study the
properties of these acoustic glitches in main-sequence stars. 
We find that the acoustic glitch due to the helium ionization zone does not
correspond to the dip in the adiabatic index $\Gamma_1$ caused by the ionization
of HeII, but to the peak in $\Gamma_1$ between the HeI and HeII ionization zones.
We find that it is easiest to study the acoustic glitch due to the helium ionization
zone in stars with masses in the range 0.9--1.2 $M_\odot$. 

\end{abstract}


\section{Introduction}
\label{sec:intro}

It is known that a steep variation in the sound speed or its derivatives inside 
a star introduces an oscillatory component, $\delta\nu$, in the frequencies of 
stellar oscillations as a function of the radial order of the eigenmodes 
\citep{goug88,voro88,goug90}, which  is proportional to 
$\sin(4\pi\tau_g\nu_{n,l}+\phi)$, where $n$, $l$, $\nu_{n,l}$, and $\tau_g$ are 
respectively the radial order,  the degree, the eigenfrequency, and the acoustic 
depth (i.e., sound travel time) of the sharp feature as measured from the stellar 
surface. These variations arise in a number of regions such as the discontinuity 
in the second derivative of the sound speed at the boundaries of the 
convection zones as well as the localized depressions in the adiabatic index $\Gamma_1$ 
in the ionization zones of abundant elements.

The important ionization zones when it comes to acoustic glitches are those where 
HI, HeI, or HeII undergo ionization. Of these, the HI ionization zone is very broad 
and the signal gets damped very quickly --- the amplitude of the signal is 
proportional to $e^{-8\pi^2\Delta^2\nu^2}$ \citep{houd07} with $\Delta$ being the 
half-width ($\sigma$) of a Gaussian profile that approximates the depression of 
$\Gamma_1$ in HI ionization zone. A typical value of $\Delta$ for 
main-sequence stars considered in this work is about 150\,s \citep{houd07}
implying that the amplitude will reduce by a factor of $e$ at a frequency of around 700\,$\mu$Hz.
Furthermore, the acoustic depth of this signal is very small 
and therefore any left-over signal behaves like a smooth function of frequency 
which makes it difficult to determine its oscillatory nature. The HeI ionization 
zone overlaps with the HI ionization zone and is again difficult to isolate. Similarly, 
the boundary of convective cores cannot be detected  because of aliasing \citep{mazu01}. 
Thus in most cases only the HeII ionization zone and the base of the envelope convection 
zone (CZ) can be probed through acoustic glitches. 

The acoustic glitches for several stars have been studied using data from {\it CoRoT} 
and {\it Kepler} missions. Using {\it CoRoT} data, \citet{migl10} determined the location of 
the second helium ionization zone for the red giant HR~7349, while \citet{roxb11} 
and \citet{mazu11,mazu12} determined the same for a solar type star HD~49933. 
\citet{mazu14} have used {\it Kepler} data to determine the depth of HeII ionization 
zone as well as the depth of the surface convection zone in 19 stars.
\citet{verm14} have used {\it Kepler} data to estimate the helium abundance of a 
binary system, 16 Cyg A and B.

In this work we study the signal expected from the HeII ionization zone and from the base of
the envelope convection zone in main-sequence stars  with masses between $0.8M_\odot$ 
and $1.5M_\odot$ using stellar models. We have restricted the study to main-sequence stars since the 
presence of mixed modes in more evolved stars make it difficult to isolate the 
oscillatory signal reliably. The mass limits are determined by the strength of 
the He signal and whether or not we expect a star to have a deep enough envelope
convection zone to excite oscillations. For stars at the subsolar mass end,
the dip in $\Gamma_1$ caused by  HeII ionization is rather shallow, and hence the amplitude of the 
oscillatory signal is very small. For stars of relatively high mass, 
greater than about 1.5$M_\odot$, the convection zone becomes 
very shallow and overlaps with the HeII ionization zone making it difficult to fit 
the signal produced by the two glitches. 
Furthermore, the envelope convection zone may split into two parts in such stars introducing 
two additional convective boundaries in a narrow region, which complicates the effective signature
from the base of the convection zone.
In this work we use stellar models in an attempt to identify stars 
for which the oscillatory signal can be reliably used to study the stellar properties.
The mass-range studied in this work is similar to \citet{basu04}, who
proposed that the acoustic glitches can be used to measure the helium abundance
in the envelope of these stars. That work was aimed at using the
amplitude of the HeII signal and did not pay much attention to the acoustic depth of the glitches.
In particular, they did not attempt to
identify the fitted acoustic depths of the acoustic glitches to specific
features in the stellar models. This issue was addressed to some extent by \citet{houd07} who
found that inclusion of acoustic glitch from HeI ionization zone improves
the agreement between the fitted acoustic depth of the glitch from HeII ionization
zone and the actual acoustic depth of the HeII ionization zone in a solar model.
In this work we wish to investigate this in more detail. 
Note that \citet{broo14} had a similar theoretical study
for the acoustic depth of the HeII ionization zone in the red giant models.

There is often a systematic offset in the acoustic depths of the glitches obtained
from fitting their signature in the frequencies with that calculated using the
sound speed profile. A part of this offset is caused by the uncertainty in the 
definition of the effective surface of the star from which the acoustic 
depth is measured. \citet{balm90} have argued that in the outer convection zone, the
squared sound speed, $c^2$,  to some approximation decreases linearly with increase in radius.
Hence, they suggested that the seismic surface can be defined as the layer at which the extrapolated 
$c^2$ vanishes. In a solar model the surface defined in this manner is located
at an acoustic height of about 225\,s above the photosphere. The uncertainty in the 
location of outer boundary affects the acoustic depths of all glitches by the same amount. 
For stellar models without overshoot, the location of acoustic glitch at the base of 
the convection zone is unambiguously defined --- this is the point where
the adiabatic temperature gradient in the convection zone changes to the
radiative temperature gradient giving rise to a 
 discontinuity in the second derivative of the sound speed. Hence we can use the fitted acoustic 
depth of the convection zone signal to estimate the location of the surface making it 
easy to compare the fitted acoustic depth of the glitches to the actual acoustic depth
in a model.
To avoid this uncertainty in definition of acoustic surface, \citet{ball04} and \citet{mazu05}
have suggested that the acoustic radius be used instead. The acoustic radius 
of a glitch is the sound travel time from the center to the location of the glitch. 
However, the form of the fitting functions involve the acoustic depths of the glitches, and 
therefore will have to be transformed to the acoustic radius using the total acoustic radius of the star.
The stellar acoustic radius is related to the large frequency separation. 
But this is not devoid of uncertainties  because of the contribution of the  surface term 
to the frequencies and hence to the large separation.  
Thus it is not clear if this transformation would help, and hence we used acoustic depth in this work.

The issue of the uncertainty in the position of the glitch is a bit more complicated 
in the case of the He ionization zones.
Unlike the convection-zone base, the ionization zones are merely regions of sharp change. They   
do not lead to a discontinuity in the derivatives of the adiabatic index, and hence, there is no discontinuity
in the derivatives of the sound speed either. 
The depressions D1, D2, and D3 in $\Gamma_1$, as shown in 
Fig.~\ref{fig:gam1}, due to HI, HeI, and HeII ionization zones respectively, result in a peak P2, 
which has a  sharper profile than the depression due to the HeII ionization. It has generally been 
assumed that the acoustic glitch whose signature we see in the frequencies is caused
by D3 where the bulk of HeII ionizes \citep{mont05,houd07}. However, the
fitted acoustic depth of the signal does not match the acoustic depth of D3
\citep{houd07,mazu14}. \citet{broo14} found that for red giant models the
fitted acoustic depth is close to that of P2. Thus it is worth investigating
which feature in the He ionization zone results in the oscillatory signature,
in particular, if it is the depression D3 as had always been assumed,
or whether it is the peak P2.
In this work we attempt to identify the location of the glitches by fitting the oscillatory signal 
to frequencies of stellar models and comparing the fitted value of $\tau_g$ with the 
acoustic depths of various features in the models.

The rest of the paper is organized as follows: Section 2 describes the techniques 
for fitting the oscillatory signal, Section 3 describes the set of stellar models 
constructed to study the acoustic glitches, Section 4 describes the results, and 
Section 5 gives a summary of results.

\section{The fitting techniques}
\label{sec:technique}

We carried out the analysis using two different techniques to fit the
oscillatory signal in the frequencies due to the major acoustic glitches
as a function of the radial order, or equivalently, the frequency.
The first technique
fits directly the frequencies whereas the second fits their second differences.
The details of the techniques are described below.  

\subsection{Fitting the frequencies directly (Method A)}
\label{subsec:freq}

We fitted the oscillation frequency, $\nu_{n,l}$, directly by modelling the smooth and 
the oscillatory components appropriately. For each degree $l$, the smooth component 
was modelled using a fourth degree polynomial in radial order $n$, and the form of the 
oscillatory signals arising from the base of convection zone and from the HeII 
ionization zone were adapted from \citet{houd07}. The full expression fitted to the 
frequency is given by
\begin{eqnarray}
f(n,l) &=& \sum_{i = 0}^{4} A_{l,i} n^{i}
              + \frac{A_c}{\nu^2} \sin(4\pi\tau_\mathrm{CZ}\nu + \psi_\mathrm{CZ}) \nonumber \\
              && + A_h \nu e^{-c_{2}\nu^2} \sin(4\pi\tau_\mathrm{He}\nu + \psi_\mathrm{He}) \,,
\label{eq:freq}
\end{eqnarray}
where $A_{l,i}$ are the coefficients of the polynomial in $n$ that defines the smooth component
of the frequencies; $A_c$ and $A_h$ give a measure of the amplitudes of the 
CZ signal and the HeII signal respectively; $c_2$ is a parameter related
to the thickness of the HeII ionization zone; $\tau_\mathrm{CZ}$ and
$\tau_\mathrm{He}$ are respectively the acoustic depths of the CZ base and the HeII ionization zone respectively;
 $\psi_\mathrm{CZ}$ and $\psi_\mathrm{He}$ define the phases of
the two oscillatory signals.
The $4\times5$ elements of $A_{l,i}$ (assuming that we are
fitting modes with degree $l$ of 0--3) along with $A_c$, $\tau_\mathrm{CZ}$,
$\psi_\mathrm{CZ}$, $A_h$, $c_2$, $\tau_\mathrm{He}$, $\psi_\mathrm{He}$
are 27 free parameters. The three terms in Eq.~(\ref{eq:freq}) are respectively 
the smooth component, signal from the base of convection zone, and signal from the
HeII ionization zone. 
It is not possible to observe $l=3$ modes in most stars, and 
therefore we use these modes only for the Sun and 16~Cyg~A, for which
they have been observed. 16~Cyg~A is one of the best-studied
stars using data from {\it Kepler} \citep{metc12, verm14} and has largest 
number of modes after the Sun with a reasonably precise set of frequencies.
For the rest of the models, we use only modes
of $l=0$--2.  Most of the results presented in this work are based on the fit to the above expression.
We have also studied for the Sun and 16~Cyg~A the possibility of separating out the oscillatory
signal caused by the HeI ionization by adding one more oscillatory term similar to
the HeII term to Eq.~(\ref{eq:freq}).

We fitted the frequencies to the function $f(n,l)$ using a nonlinear least-squares 
fit with second derivative smoothing. The smoothing, which is applied
to only the first term in Eq.~(\ref{eq:freq}), provides additional
constraints that enable us to determine a relatively large number of parameters.
We used the same value of the smoothing 
parameter as in \citet{verm14}. Since the nonlinear minimization may not converge to the 
global minimum for different starting guesses, we repeated the minimization with 
multiple sets of initial guesses (100 when fitting only the HeII and CZ signals, 500 
when the HeI term is added)
of the free  parameters. The different sets of initial guesses were
obtained by randomly perturbing a reasonable value for each parameter.
The solution with the minimum among the set of $\chi^2$ 
generated in these trials was accepted as the best fit to the data. 
In order 
to estimate the uncertainties on the fitted parameters, the fitting process 
was repeated for 1000 realizations of the data obtained by perturbing 
the frequencies with Gaussian random errors with standard deviation equal 
to the uncertainties in the  frequencies. The uncertainties
used depends on the star or model being fit and is discussed
further in Section~\ref{sec:results}.

\subsection{Fitting second differences (Method B)}
\label{subsec:diff}

In this technique, we enhanced the oscillatory signal by taking the second 
differences \citep{goug90,basu94,basu04,mazu05} of the frequencies with respect to the radial 
order, $n$,
\begin{equation}
\delta^2\nu_{n,l}=\nu_{n-1,l}-2\nu_{n,l}+\nu_{n+1,l}.
\label{eq:dif}
\end{equation}
The main advantage of taking second differences is that it removes the contribution 
from the dominant smooth trend which is a linear function of $n$. On the other hand, 
taking the second differences introduces correlations between neighboring points 
that need to be accounted for by using the covariance matrix while defining the 
$\chi^2$ function. Another disadvantage of this technique is that the amplitude 
of oscillatory signal from small $\tau_g$ features is reduced significantly. We 
fitted the second differences to oscillatory signals from the base of the convection 
zone and the HeII ionization zone \citep{mazu01}. We used the following form which 
has been adapted from \citet{houd07},
\begin{eqnarray}
\delta^2\nu&=& a_0+a_1\nu+a_2\nu^2+{b_2\over\nu^2}\sin(4\pi\nu\tau_\mathrm{CZ}+\phi_\mathrm{CZ})\nonumber\\
&&\qquad + c_0\nu e^{-c_2\nu^2}\sin(4\pi\nu\tau_\mathrm{He}+\phi_\mathrm{He})\,,
\label{eq:sd}
\end{eqnarray}
where $a_0, a_1, a_2, b_2, c_0, c_2, \tau_\mathrm{CZ}, \tau_\mathrm{He}, \phi_\mathrm{CZ}, 
\phi_\mathrm{He}$ are 10 free parameters. The first three terms represent the smooth 
part of the function which remains after the second differences are calculated, the 
next term represents the contribution from the base of the convection zone
with $b_2$ giving a measure of the amplitude of the signal, $\tau_\mathrm{CZ}$ its
acoustic depth, and $\phi_\mathrm{CZ}$ the phase,
while the 
last term represents the contribution from the HeII ionization zone
with $c_2$ related to the thickness of the ionization zone, 
$c_0$ giving a measure of the amplitude of the signal, $\tau_\mathrm{He}$ its
acoustic depth, and $\phi_\mathrm{He}$ the phase.
The number of 
terms needed to approximate the smooth part depends on the range of frequencies that are 
observed and the errors in these frequencies.
Since in this case we are mainly dealing with model frequencies, we use the 
observed solar frequencies to decide the number of terms needed. Adding more terms
doesn't lead to a statistically significant reduction in $\chi^2$. Since this method 
uses fewer parameters as compared to Method A, in some cases the fit is 
not as good as that for Method A. This is particularly the case when the frequency 
range used is large and the errors in frequencies are small, as is the case for the 
Sun and for frequencies of stellar models. An important difference between the two 
methods is that in Method B, the smooth part is independent of $l$, which may be 
justified because a large part of the smooth trend gets filtered out when second differences 
are taken. But in some cases a residual $l$ dependence may remain in the second 
differences also. We also repeated the exercise after including an additional term for 
HeI ionization zone for the Sun and 16 Cyg A, as was done in Method A.

The parameters in Eq.~(\ref{eq:sd}) were determined using a nonlinear least-squares 
fit, with $\chi^2$ defined using the covariance matrix. The covariance matrix was 
calculated by assuming that the errors in individual frequencies are not correlated. 
This is the usual approximation when using solar and stellar frequencies.
Although, all frequencies are determined by using a single oscillation
power spectrum, the frequencies are reasonably well separated and
the correlation between frequencies of two different modes is quite small.
If the covariance matrix for the observed frequencies is available it can be easily taken into
account while calculating the error covariance matrix for the second differences.
As in Method A, we made multiple (100/500) attempts to fit the signal
using different initial guesses for the 
free  parameters, which were obtained by randomly perturbing a
reasonable value of the initial guesses.
The minimum of the set of $\chi^2$ obtained over these multiple attempts was chosen to be 
the best fit. 
Similarly, the errors in the fitted parameters were estimated by repeating the 
whole process for 1000 realizations of data obtained by randomly perturbing
the frequencies.

\section{Stellar models Used}
\label{sec:model}

The models were constructed using the evolutionary code MESA \citep{paxt11}. We 
used OPAL equation of state \citep{roge02}, OP high temperature opacities 
\citep{badn05,seat05} supplemented with low temperature opacities from \citet{ferg05}. 
The metallicity mixture of \citet{gs98} was used. We used the reaction rates from 
NACRE \citep{angu99} for all reactions except $^{14}$N($p$,$\gamma$)$^{15}$O and 
$^{12}$C($\alpha$,$\gamma$)$^{16}$O, for which updated reaction rates from
\citet{imbr05} and \citet{kunz02} were used. Convection was modelled using the 
standard mixing length theory \citep{cox68} without overshoot, and diffusion of helium and heavy 
elements was incorporated using the prescription of \citet{thou94}.

\begin{figure}
\plotone{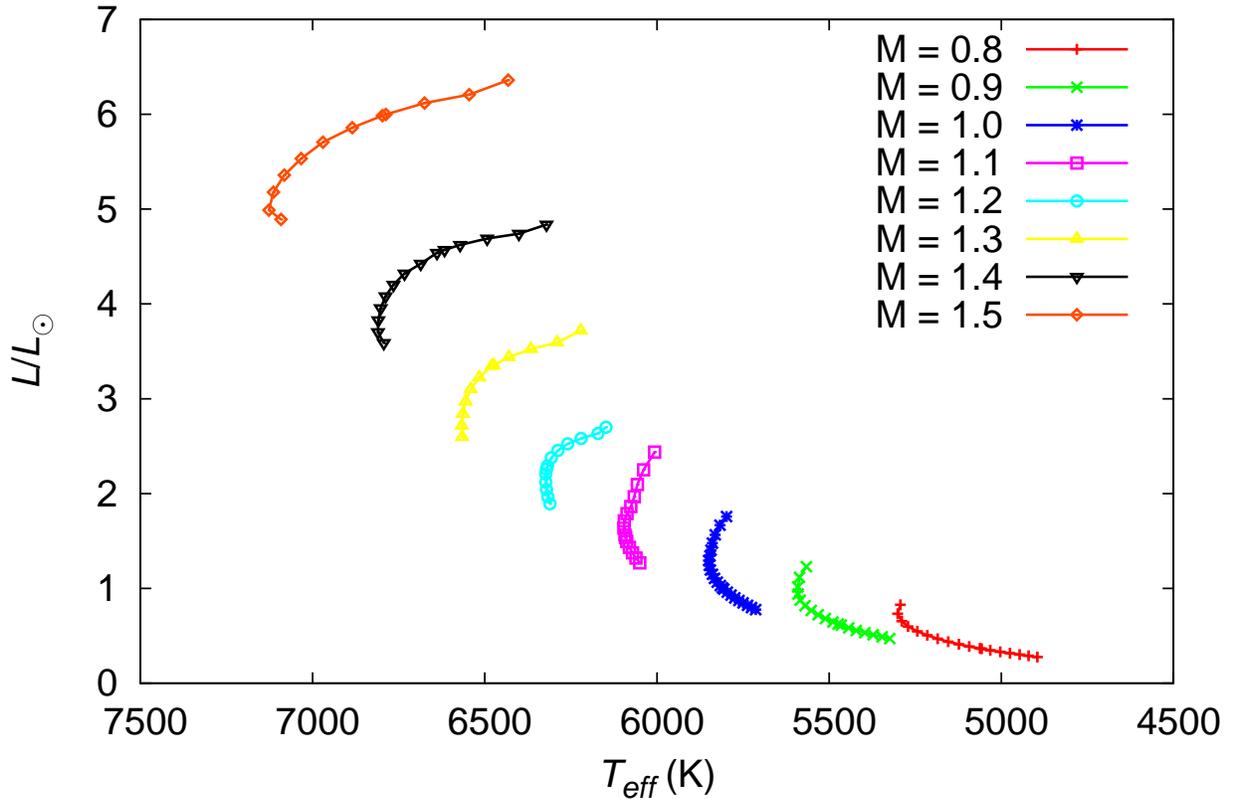}
\caption{H-R diagram showing the evolutionary stages of the models under study.}
\label{fig:hr}
\end{figure}

We constructed models with an initial helium abundance of $Y_i = 0.28$, initial heavy element 
abundance of $Z_i = 0.02$, and mixing length parameter $\alpha = 1.91$ as obtained 
from solar calibration. The models cover a range in mass and age. The envelope helium 
and heavy element abundance get depleted for models of mass greater than $1.4M_\odot$ 
because of diffusion, therefore for these masses we use models without diffusion.
In these cases we use  a mixing-length parameter of  $\alpha = 1.84$ as is obtained for a
calibrated  solar model without diffusion. Fig.~\ref{fig:hr} shows the evolutionary 
stages of the models in the H-R diagram. 

In addition to these evolutionary sequences, we also constructed one solar model and 
a representative model for the solar analog 16~Cyg~A. The 16~Cyg~A representative model 
was constructed with a mass of $1.05M_\odot$, an initial helium abundance of 0.29, 
initial heavy element abundance 0.022, and age 6.9 Gyr. 

\section{Results}
\label{sec:results}

Before looking at models of other stars where there could be uncertainties in
radius and luminosity, we first examined what happens if we compare results
from a solar model and the Sun. 
Since the mass, radius, and luminosity of the Sun
are known independently, solar models have the same radius and luminosity 
as the Sun. We also fitted the frequencies of 16~Cyg~A and its model.

The uncertainties in the input frequencies affect the nature of the fits
and determine the uncertainties in the fitted parameters.
To estimate this uncertainty, we repeated the fitting process for each star/model
for 1000 realizations of the data obtained by perturbing the frequencies with Gaussian
random error with standard deviation equal to the quoted error-bars/weights on the
frequencies. Since the evolutionary sequence of models are generic models
and not those of any particular star with observed frequency estimate, there is no 
statistical uncertainty in the frequencies and the fitted parameters. However, we 
assumed that the frequencies of the models of the Sun and 16~Cyg~A have the
same weights as the corresponding observed frequency estimate. This was done to
 avoid systematic errors in the fitting.
For other stellar models we assume an error of 0.1 $\mu$Hz in all modes,
which is a reasonable estimate of
the uncertainties in the asteroseismic data, and therefore gives an idea of the 
precision to which we can determine the glitch parameters of a real star.

\subsection{Results for the Sun and 16~Cyg~A}

We used the solar data set obtained by the Birmingham
Solar Oscillation Network \citep[BiSON;][]{elsw91} listed in
Table~1 of \citet{chap07}. The frequencies of 16~Cyg~A were those obtained
by NASA's {\it Kepler} mission and listed in Table~2 of \citet{verm14}.
\citet{vern06} had shown that it is possible to determine the signature of acoustic glitches 
in low-degree modes obtained by BiSON if the data covers an interval of 6 months or longer, and thus
we were confident that we would be able to fit the signatures of the acoustic glitch in the
frequencies obtained from a much longer time-series.
We also fitted the frequencies of models of the Sun and 16~Cyg~A.
We determined the parameters in Eqs.~(\ref{eq:freq}) and (\ref{eq:sd}) using the techniques described in 
Section~\ref{sec:technique}.
The modes of degree 0, 1, 2, and 3 were used in the fits for the Sun (total 72 modes in 
the frequency range of 1.4-4.0 mHz) and 16~Cyg~A (total 53 modes in the frequency range of 1.3-2.9 mHz).

\subsubsection{Fitting only the HeII and CZ signals}

The fits to the BiSON data are shown in Fig.~\ref{fig:fit}.
The left panel of Fig.~\ref{fig:fit} shows the result of fitting
Eq.~(\ref{eq:freq}) to the observed frequencies using Method A.
To see the oscillatory component, $\delta\nu$, clearly, we subtracted the 
smooth component fitted by the polynomial in Eq.~(\ref{eq:freq}).
The lower panel shows the normalized residuals of the fit
obtained by dividing the residual of the fit by the error in the frequency. Note 
the significantly large residual and the oscillatory trend in it. This could be due to inaccurate 
modelling of the oscillatory signal, and will be explored later. The fitted 
parameters obtained using Methods A and B are listed in Table~\ref{tab:par1} for 
both the Sun and a solar model. In the table, $\chi^2$ denotes the weighted 
least-squares residual of the fit, $A_\mathrm{CZ}$ and $A_\mathrm{He}$ denote 
respectively the amplitude of CZ and He signal averaged over the frequency range 
used in the fit, and $\Delta_\mathrm{He}$ denotes the half-width of the glitch as 
obtained using fitting parameter $c_2$ of Eqs.~(\ref{eq:freq}) and (\ref{eq:sd}) 
($\Delta_\mathrm{He} = \sqrt{c_2/8\pi^2}$). 
The amplitudes of the signals obtained by fitting the
second differences in  Method B have been converted to the amplitude of the
signal in the frequencies
by dividing the second-difference amplitudes by $4\sin^2(2\pi\tau_g\Delta_0)$, 
where $\Delta_0$ is the large frequency separation \citep{houd07}.

\begin{figure}
\plotone{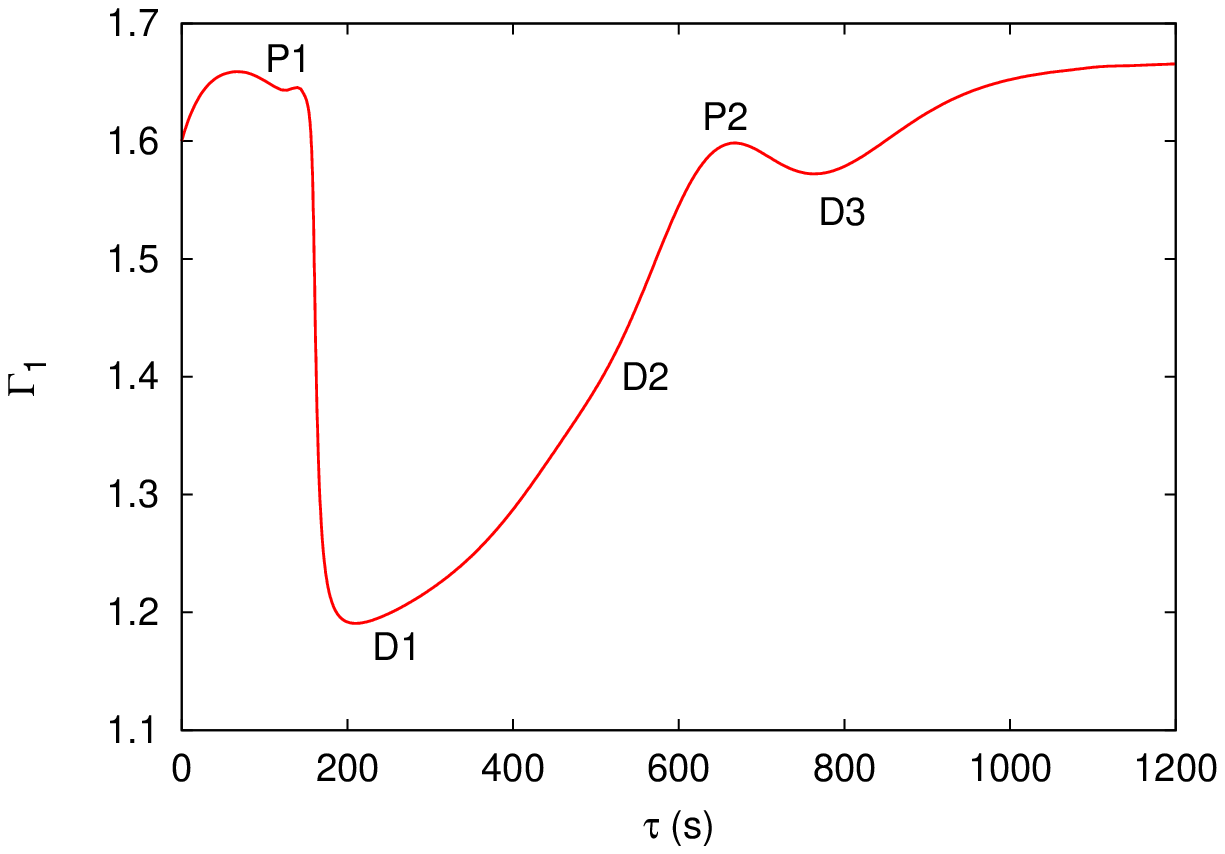}
\caption{The first adiabatic index as a function of acoustic depth of a solar model.
The labels D1, D2, and D3 refer to the hydrogen ionization zone, first helium ionization 
zone, and second helium ionization zone respectively, while P2 refers to
peaks in $\Gamma_1$ that arise between the HeI and HeII ionization zones. On the
other hand P1 is above the HI ionization zone.}
\label{fig:gam1}
\end{figure}

\begin{figure}
\plottwo{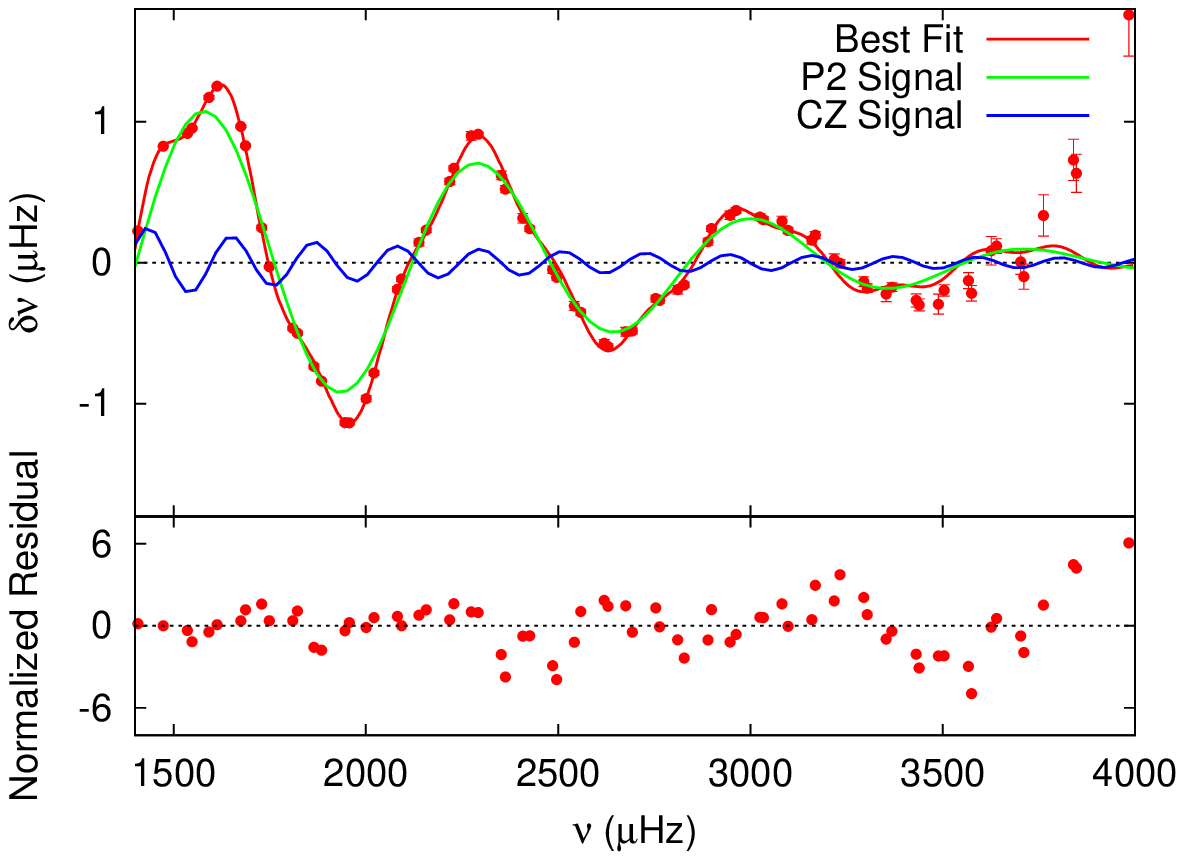}{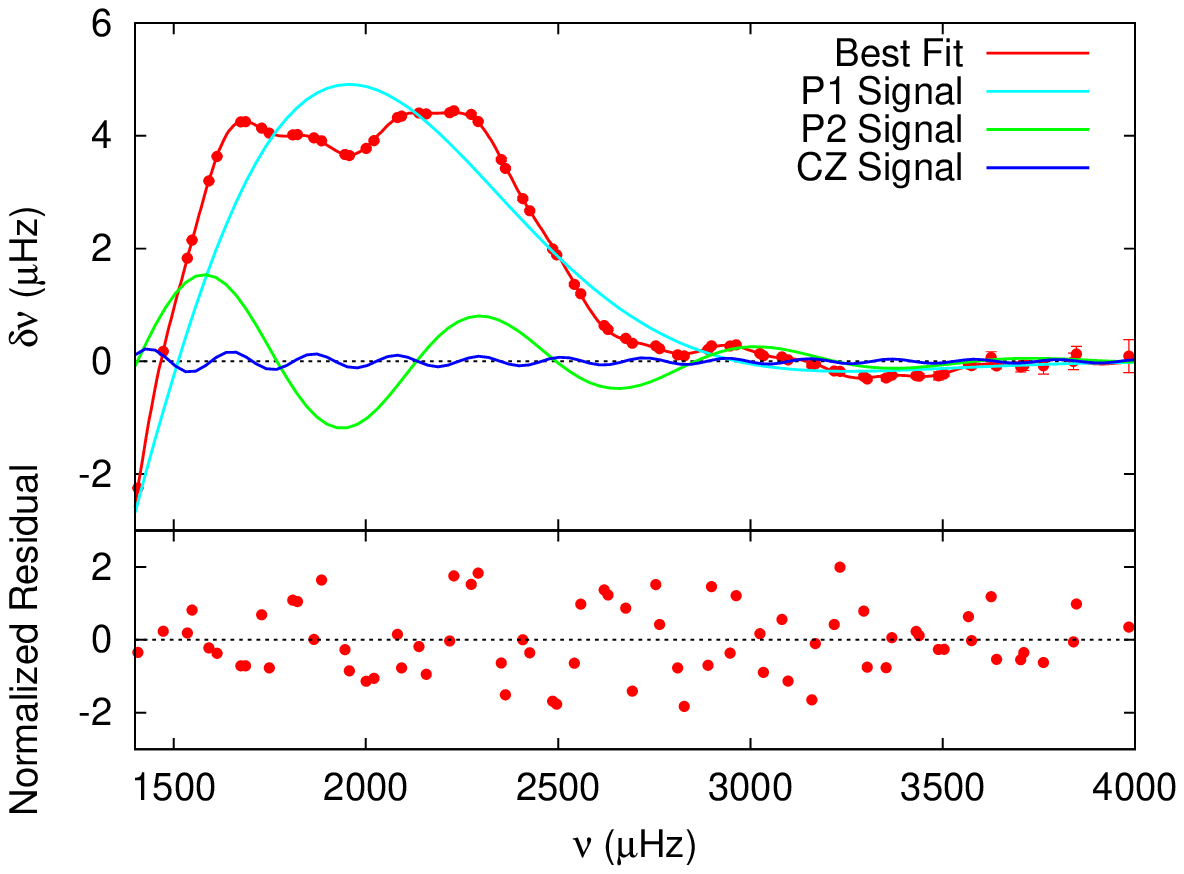}
\caption{The fit to the observed BiSON frequencies using Method A.
The figure shows the oscillatory part of the frequency ($\delta\nu$) obtained
by subtracting the smooth part from the frequencies. The dots show the observed
BiSON frequencies. The left panel 
shows the oscillatory component of the fit ($\delta\nu$) when only one
glitch from He ionization zone is included. The lower panels show the normalized residuals, which are obtained by 
dividing the residual with the error on the frequency. The right panel shows the 
fit when the oscillatory signal due to P1 (see Fig.~\ref{fig:gam1}) is also included 
in the fit.}
\label{fig:fit}
\end{figure}

From Table~\ref{tab:par1} we can see that the fitted $\tau_\mathrm{He}$ and 
$\tau_\mathrm{CZ}$ for the solar model and the Sun are in close agreement with each 
other with the difference being about 1\%. However, in the case of the solar model we find that the fitted value of
$\tau_\mathrm{CZ}$ is larger than the acoustic depth of the base of convection 
zone (2244~s) as measured from the top of the atmosphere at an optical depth of 
$10^{-5}$. This difference can be attributed to the choice of the stellar surface and 
by extending the model further by about 400 km it is possible to match the two values. 
We find that if the model is extended to an optical depth of $2\times10^{-6}$, the
fit does not change much, for example the fitted $\tau_\mathrm{CZ}$ using Method A 
is $2294\pm5$~s, while the calculated acoustic depth is 2301~s. Clearly both of them are 
now in good agreement with each other. The issue is different with $\tau_\mathrm{He}$. The fitted 
value of $\tau_\mathrm{He}$ is smaller than the calculated acoustic depth of the 
HeII ionization zone defined as the  minimum in $\Gamma_1$ (D3 in Fig.~\ref{fig:gam1}), 
the calculated acoustic depth of this dip is $\tau_\mathrm{D3}= 764$~s. The fitted 
value is in fact close to the acoustic depth of a layer above the HeII ionization 
zone where $\Gamma_1$ is maximum (P2 in Fig.~\ref{fig:gam1}) which has a calculated 
acoustic depth of $\tau_\mathrm{P2} = 668$~s. Extending the stellar atmosphere 
upwards further increases the acoustic distance between the fitted $\tau_\mathrm{He}$ 
and $\tau_\mathrm{D3}$, while the fitted $\tau_\mathrm{He}$ remains consistent with 
$\tau_\mathrm{P2}$.

We repeated the same exercise as above for 16 Cyg A and the results are listed in 
Table~\ref{tab:par1}.
The table also shows the results for a representative stellar model for this star.
Similar to the solar model, the results for the model of 16~Cyg~A also
shows that the fitted $\tau_\mathrm{He}$ is closer to the acoustic depth of 
point P2 (911~s) rather than D3 (1051~s).

\subsubsection{Fitting the signal from HeI ionization zone explicitly}

\citet{houd07} have argued that there should be an oscillatory contribution to the 
frequency from HeI ionization zone too; however, it is not clear whether the 
contribution is significant and detectable since the HeI ionization zone overlaps with HI 
ionization zone. The residuals shown at the bottom of the left panel of Fig.~\ref{fig:fit} seem to 
have an oscillatory signal of period about 500 $\mu$Hz, which corresponds to a 
glitch at an acoustic depth of 1000~s. This acoustic depth does not
correspond to the (shallow) HeI ionization zone, but to a layer 
just {\it below} the HeII ionization zone where $\Gamma_1$ is close to
its asymptotic value of 5/3
(see Fig.~\ref{fig:gam1}). 

To study the  effect of the HeI ionization zone we included one more term, similar to the last term, in 
Eq.~(\ref{eq:freq}) resulting in four additional parameters. Two mathematically similar terms 
usually destabilize a fit, and hence, to stabilize the fit we fixed the ratio of the acoustic depths of HeI and HeII 
ionization zones ($\eta \equiv \tau_\mathrm{I}/\tau_\mathrm{II}$) and keep the other 
three parameters free. This differs from the methodology of \citet{houd07} who fixed 
all four parameters using theoretically expected ratios between the parameters of the
HeI and HeII signals. We varied the ratio $\eta$ in the range 0.2--0.9 and obtained the 
best fit to the data for each value of $\eta$. The fitted parameters obtained using 
Method A for the solar model and the BiSON frequencies are listed in Table~\ref{tab:par2} 
for different values of $\eta$. Clearly the two helium-glitch model improves the fit with 
$\chi^2$ reducing significantly compared to those in the Table~\ref{tab:par1}. 
However, the $\chi^2$ has a minimum at around 
$\eta = 0.25$, and the corresponding fit to the observed frequencies is shown in 
the right panel of Fig.~\ref{fig:fit}. Since Method A fits the smooth and oscillatory
components of frequency together, the addition of the second helium-glitch changes the
smooth component as well. As a result, the oscillatory signal $\delta\nu$ obtained by
subtracting the smooth part of frequency, looks very different in the two
panels. The fit corresponds to the helium-glitches at 
acoustic depths of 172 s and 689 s for the observed frequencies, and at 170 s and 
678 s for the solar model. These acoustic depths do not correspond to the HeI 
and HeII ionization zones, instead, one of them again corresponds to the peak P2 
in $\Gamma_1$ profile and the other corresponds to a peak P1 near the surface 
(see Fig.~\ref{fig:gam1}).
Note that the fitted values of both, $\tau_\mathrm{He}$ and the amplitude 
of the He signal, approaches that obtained in Table~\ref{tab:par1} as $\eta$ approaches
0.25 where $\chi^2$ is minimum. For  smaller value of $\eta$, the $\chi^2$ of the fit
increases, the fit becomes unstable, and it is difficult to determine the
value of any particular parameter reliably.

\begin{figure}
\plotone{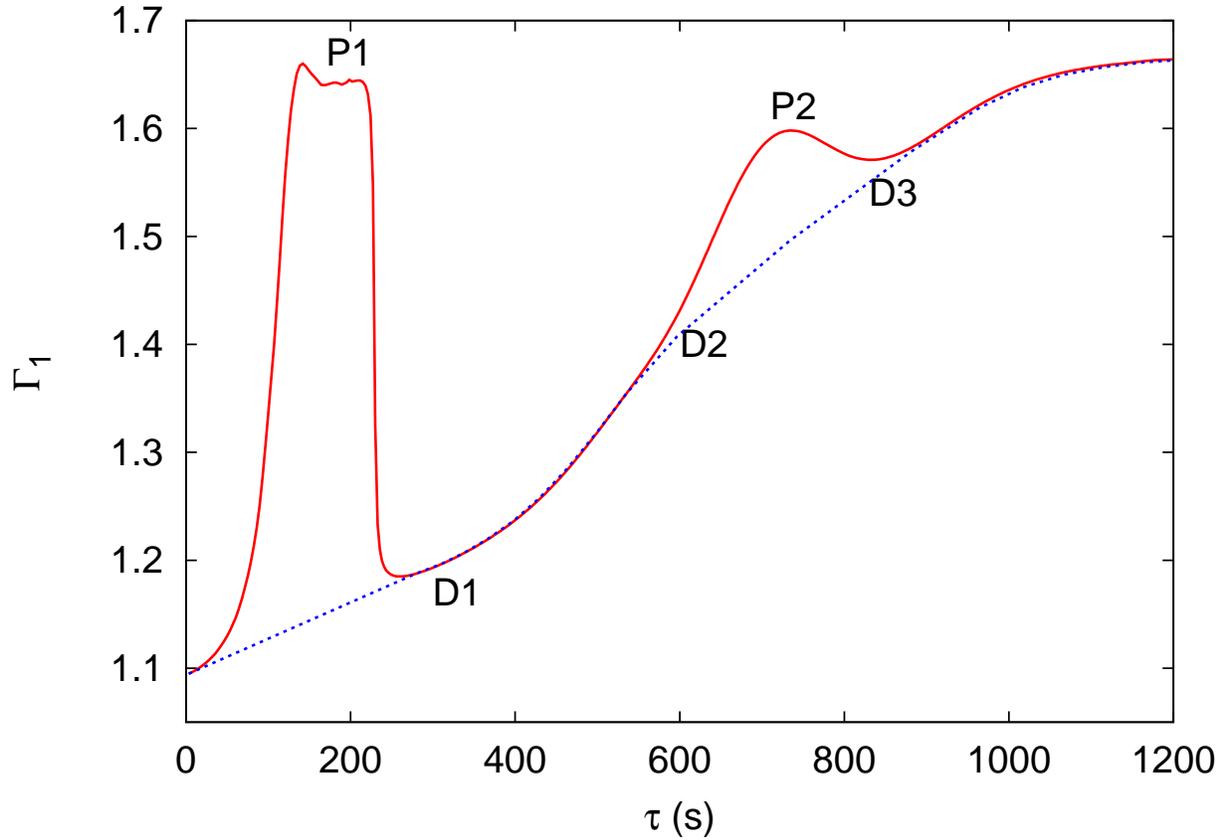}
\caption{The first adiabatic index as a function of acoustic depth of a solar model constructed 
with a realistic atmosphere. The atmosphere extends to an optical depth of $2\times10^{-6}$. 
The dotted line is an eye guide to the $\Gamma_1$ profile that contributes to the 
smooth component of the frequency.}
\label{fig:gam11}
\end{figure}

It may be noted that in the solar atmospheric model constructed with the MESA code,
the temperature asymptotically approaches a constant value and hydrogen is not 
ionized. On the other hand, the temperature increases with height beyond 
the temperature minimum in the Sun, and hydrogen gets ionized once again, giving another dip 
in $\Gamma_1$ in the atmosphere, which results in the peak P1. Fig.~\ref{fig:gam11} 
shows the $\Gamma_1$ profile in a solar model where the atmospheric model of 
\citet{vern81} is added at the top. In this atmospheric model the temperature 
increases beyond a height of about 500 km and HI gets ionized giving another dip 
in $\Gamma_1$.

The above analysis suggests that the total frequency of a solar-like star can effectively be written as, 
$\nu = \nu_s + \delta\nu_\mathrm{I} + \delta\nu_\mathrm{II} + \delta\nu_{\mathrm{CZ}}$, 
where $\nu_s$ is the smooth part of the frequency coming from the smooth profile of 
$\Gamma_1$ as shown by the dotted line in Fig.~\ref{fig:gam11}, $\delta\nu_\mathrm{I}$ and 
$\delta\nu_\mathrm{II}$ are the contribution of the peaks on top of the dotted line, and 
$\delta\nu_\mathrm{CZ}$ is the contribution of the glitch at the base of the convection zone.

Similar results were obtained by fitting the second differences using Method B. In this 
case, the $\chi^2$ is much larger and the reduction on adding the HeI term is rather 
modest, but $\chi^2$ still decreases with $\eta$. For the solar model, $\chi^2$ decreases 
from a value of 1124 when the HeI term is not included to 988 for $\eta=0.8$ and to 976 
for $\eta=0.2$. Similarly, for the observed frequencies the $\chi^2$ reduces from
a value of 1080 without HeI term to 1009 for $\eta=0.8$ and 1001 for $\eta=0.2$.
This behavior may be expected since taking the second difference modifies the 
amplitude of the oscillatory signal by a factor of $4\sin^2(2\pi\tau_g\Delta_0)$, 
where $\tau_g$ is the acoustic depth of the glitch and $\Delta_0$ is the large frequency 
separation. This factor is 1.23 for the HeII signal, while it reduces to 0.82, 0.48, 
0.22 and 0.06 for the HeI signal when $\eta = 0.8$, 0.6, 0.4 and 0.2 respectively. 
However, the errorbars in the second differences increase by about a factor of 2.5 
as compared to those in the frequencies. As a result, this method is not effective 
in detecting the oscillatory signal from glitches at low acoustic depths and the improvement 
is not as great as in the case of Method A when additional term is included in the fit. 

Since the acoustic depth of the near-surface glitch (from P1) is very small, if the frequency 
range included in the fit is not large enough to show the oscillatory signal, its 
contribution will appear as a smooth component of the frequency. This is particularly 
true if we do not have a sufficient number of low-frequency modes. To check whether 
we can detect the signal from the near-surface glitch for stars other than the Sun,
we repeated the exercise  above for 16 Cyg A
using Method A, with and without the additional HeI term to see whether it improves the fit. 
Note from Table~\ref{tab:par1} the small values of the $\chi^2$ for 
the model frequencies, which indicates that the one helium-glitch model is very close to 
the true model in the observed frequency range of this star.
We found that the additional term does not improve the 
fit -- the $\chi^2$ does not reduce significantly irrespective of the value of $\eta$.
For example, the $\chi^2$ for the observed frequencies of 16 Cyg A
reduces to 72.0 when $\eta=0.25$.
Similar results were found when we fitted the observed solar frequencies in a 
frequency range restricted by the radial orders of the available 16 Cyg A data. This suggests 
that the observed frequency range of 16 Cyg A is not large enough to detect the 
signal from the near-surface glitch.

\subsubsection{Fitting the HeI signal using artificial model}

\begin{figure}
\plotone{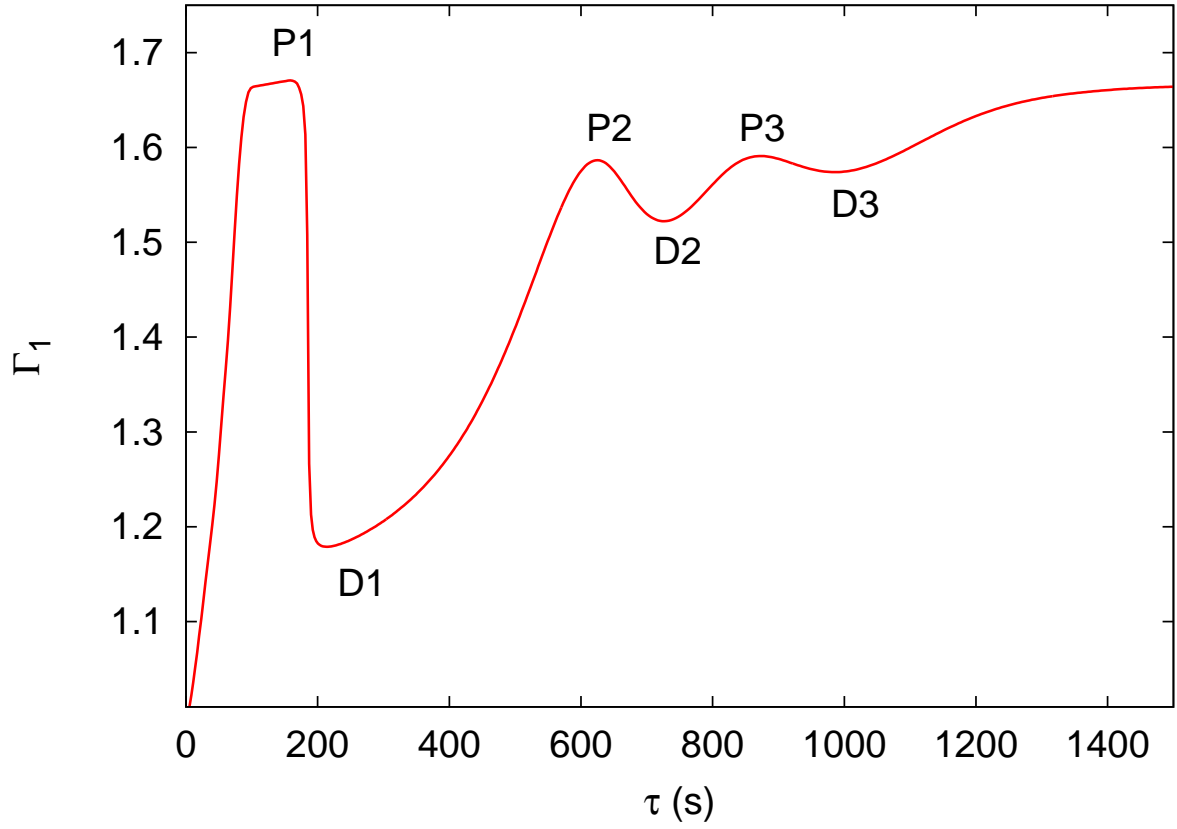}
\caption{The first adiabatic index as a function of acoustic depth for a solar
model constructed with increased helium ionization potentials to separate out 
the HI and HeI ionization zones.}
\label{fig:gam2}
\end{figure}

To check if we can expect to detect the presence of the oscillatory signal from the 
HeI ionization zone at all, we constructed a solar model with artificially increased helium  
ionization potentials to 54 eV for HeI  and 94 eV for HeII. 
This allows us to separate out the HI and HeI ionization zones. 
The $\Gamma_1$ profile for this model is shown in Fig.~\ref{fig:gam2}. To suppress
the signal from P1, we exclude frequencies at the lower end and fit the signal in the
frequency range 1.9--4.0 mHz using Method A. A single helium-glitch model fits the oscillatory signal at 
$\tau_\mathrm{He} = 687$ s with $\chi^2 = 47.1$. The fitted $\tau_\mathrm{He}$ in this case
is closer to the peak P2 between the HI and the HeI zones as that peak is sharper
than the peak P3.  Including an additional oscillatory term as above reduced the 
$\chi^2$ to 10.6  at  $\eta = 0.65$. In other words, we definitely
fit signals  from both the HeI and HeII ionization zones, i.e., if the HI 
and HeI ionization zones are separate, we can isolate the HeI signature. The 
fitted acoustic depths were found to be 620 s and 949 s, both of which correspond 
to the peaks in $\Gamma_1$ just above the respective ionization zones (labelled as 
P2 and P3 in Fig.~\ref{fig:gam2}). Thus it is clear that for the cases considered 
thus far the fitted acoustic depths $\tau_\mathrm{He}$ correspond to the peaks in $\Gamma_1$ 
above the ionization zones and not the dips in $\Gamma_1$ caused by the process of 
ionization. Similar result was found by \citet{broo14} for red giant stars.

In the limited frequency range that is observed for stars other than the Sun, the 
contribution of the near-surface glitch cannot be separated from the smooth component 
of the frequency, and is thus difficult to fit. However it may be noted from 
Tables~\ref{tab:par1} and \ref{tab:par2} that the parameters of the glitch between HeI and HeII
ionization zone do not depend on whether one helium-glitch model is fitted or two helium-glitch 
model with $\eta = 0.25$ is fitted. Therefore, we can reliably study the properties of the 
peak in the $\Gamma_1$-profile (P2 in Fig.~\ref{fig:gam11}) by fitting the frequencies 
to a single glitch from ionization zones. Hence all results in the next subsection 
are obtained using such fits. It may be noted that this single glitch
corresponds to the peak between the HeI and HeII ionization zones and hence
we refer to it as ``due to He ionization zones''.

\subsection{The He and CZ signals of other main-sequence models}

We fitted the frequencies, and the second differences of frequencies, of all the 
models described in Section~\ref{sec:model} to Eqs.~(\ref{eq:freq}) and (\ref{eq:sd}).
The fits used 48 modes around the frequency of maximum power, $\nu_{\rm max}$, which 
was calculated using the usual scaling relation \citep{kjel95}. We used only low-degree modes 
$l = 0$, 1, 2 that we expect to observe on most stars.
As mentioned earlier, we assumed a nominal
uncertainty of 0.1 $\mu$Hz for each mode for the purpose of defining the weights in the fits. 

\begin{figure}
\epsscale{0.9}
\plotone{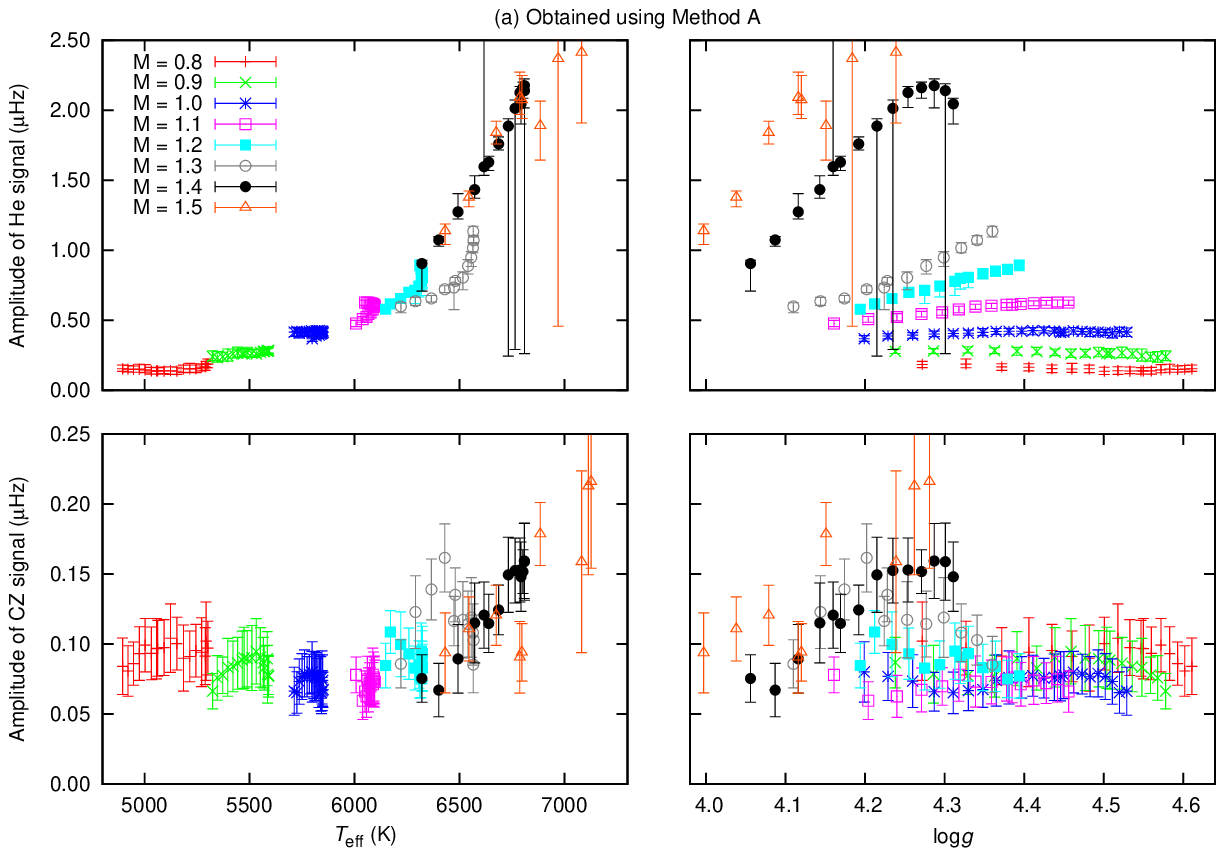}
\plotone{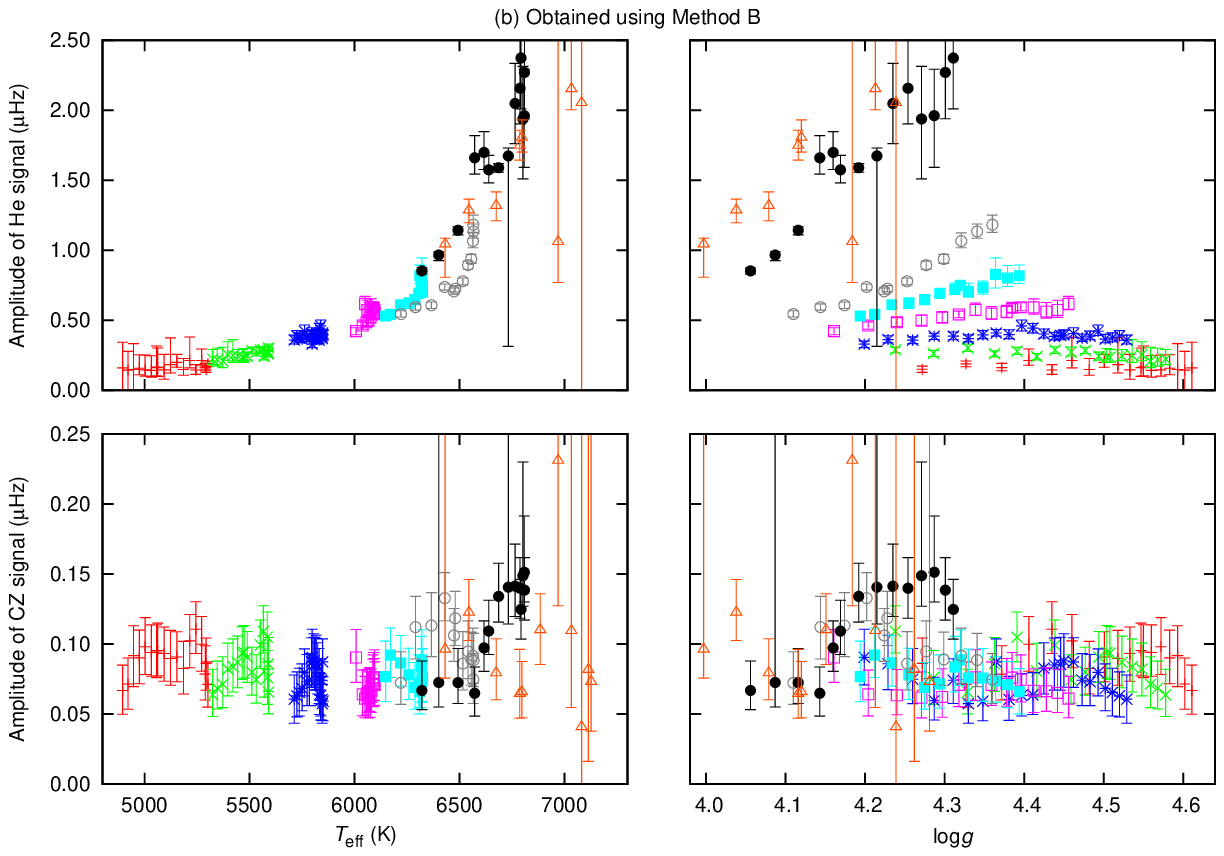}
\caption{The amplitude of He and CZ signal averaged over the frequency interval
used in the fit, as a function of the effective temperature and the logarithm of 
the surface gravity. Different types of points correspond to the masses of the 
stars as shown in top left panel.}
\label{fig:amp}
\end{figure}

Fig.~\ref{fig:amp} shows 
the fitted average amplitude of He and CZ signal as a function of effective temperature  
($T_\mathrm{eff}$) and logarithm of the surface gravity ($\log{g}$). 
The amplitude of He signal increases with effective temperature. The change with
$\log g$ at higher masses is a reflection of the change in $T_\mathrm{eff}$ as
the star evolves. It may be noted that the models of mass $1.4M_\odot$ and $1.5M_\odot$ were
constructed without diffusion, and therefore are physically different from the rest of the 
models and fall slightly off the trend in the figure.
For a star of given mass and/or effective temperature, the amplitude of signal due to He 
ionization zone depends on the amount of helium present there, hence it can be used 
to determine the helium abundance \citep{basu04,mont05,verm14}. To determine the 
helium abundance, the amplitude of the helium signal can be calibrated with the 
models of similar mass and effective temperature with different helium abundance 
to estimate the current envelope helium abundance of the star. The amplitude of 
CZ signal seems to have a minimum around $T_\mathrm{eff}=6000$~K. The increase
in amplitude as $T_\mathrm{eff}$ reduces is quite modest, but for higher
values of $T_\mathrm{eff}$ the amplitude increases more rapidly.

\begin{figure}
\plotone{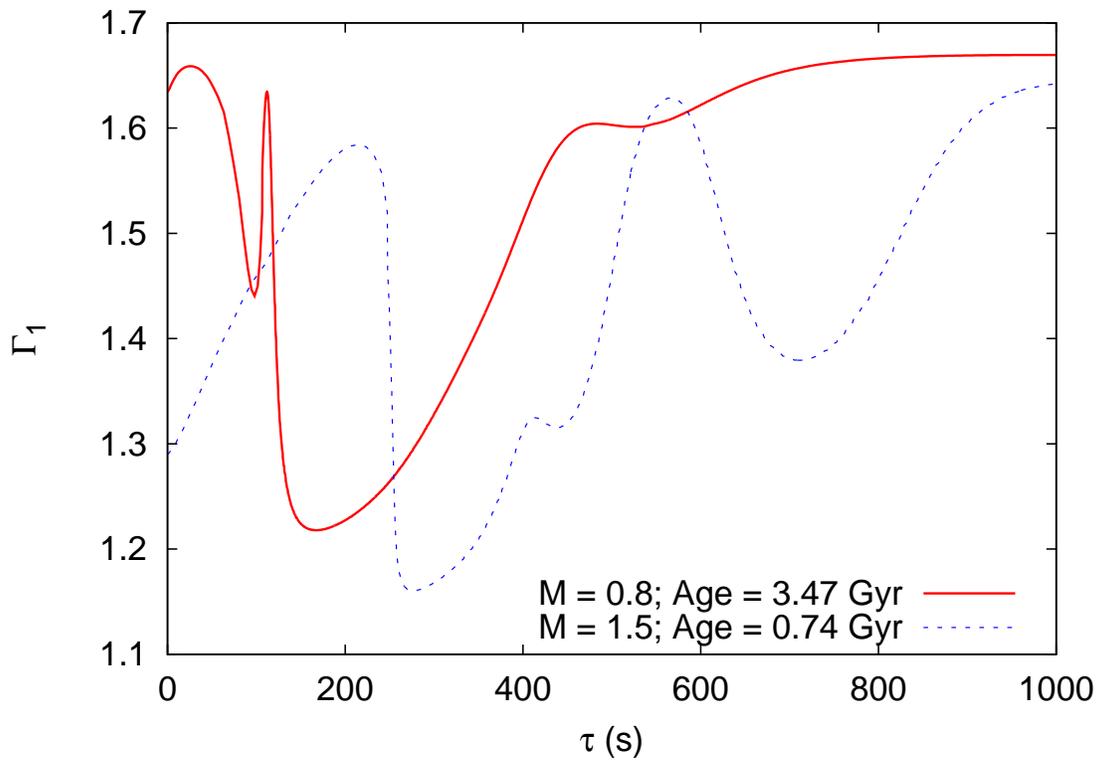}
\caption{The typical profiles of first adiabatic index for models of masses $0.8M_{\odot}$
and $1.5M_\odot$.}
\label{fig:gam815}
\end{figure}

Fig.~\ref{fig:gam815} shows the $\Gamma_1$ profiles for typical stellar models of mass 
$0.8M_\odot$ and $1.5M_\odot$ with roughly the same helium abundance. It can be seen that 
for low mass stars the dip in $\Gamma_1$ due to HeII ionization zone is very shallow, 
which reflects in the small amplitude of the oscillatory signal and can be fitted only 
if low frequency modes are included. The small peak around $\tau_g = 180$ s is due to the 
transition between the interior and the atmospheric model in the stellar model. Models with
lower masses have even shallower dips making their signal almost impossible to fit. For 
$1.5M_\odot$ stellar model the dip in $\Gamma_1$ due to HeII ionization zone is very pronounced
and even the kink due to HeI ionization zone is visible. 
But the fits for higher mass stellar models
are difficult because the HeII ionization zone and the base of the convection
zone are relatively close, confusing their signal. 
Furthermore, these stars have large composition gradient at the boundary of the shrinking 
convective core, which results in a strong peak in the Brunt-V\"ais\"al\"a frequency, $N$. 
This introduces additional effects that are not modelled by the fitting function used. 
Similar peaks in the Brunt-V\"ais\"al\"a frequency may also be seen in the lower mass stars
at the end of their main-sequence life due to large composition gradient in the core. 

\begin{figure}
\plotone{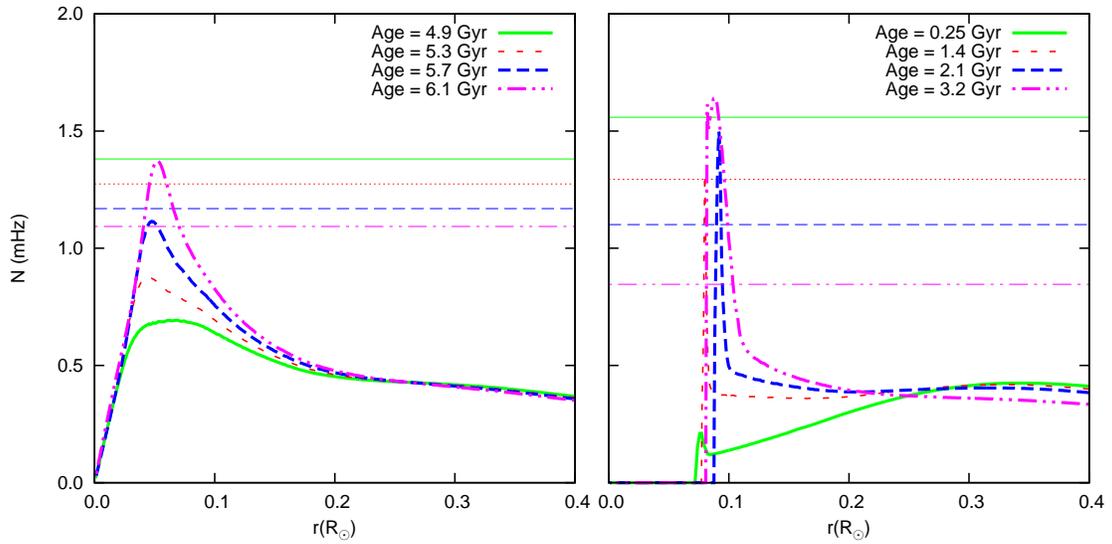}
\caption{The Brunt-V\"ais\"al\"a frequency of selected models of mass $1.1M_\odot$ 
(left panel) and $1.3M_\odot$ (right panel) as a function of radial coordinate. The 
horizontal lines correspond to the lowest frequency used in fitting the signal.}
\label{fig:bv}
\end{figure}

\begin{figure}
\epsscale{0.85}
\plotone{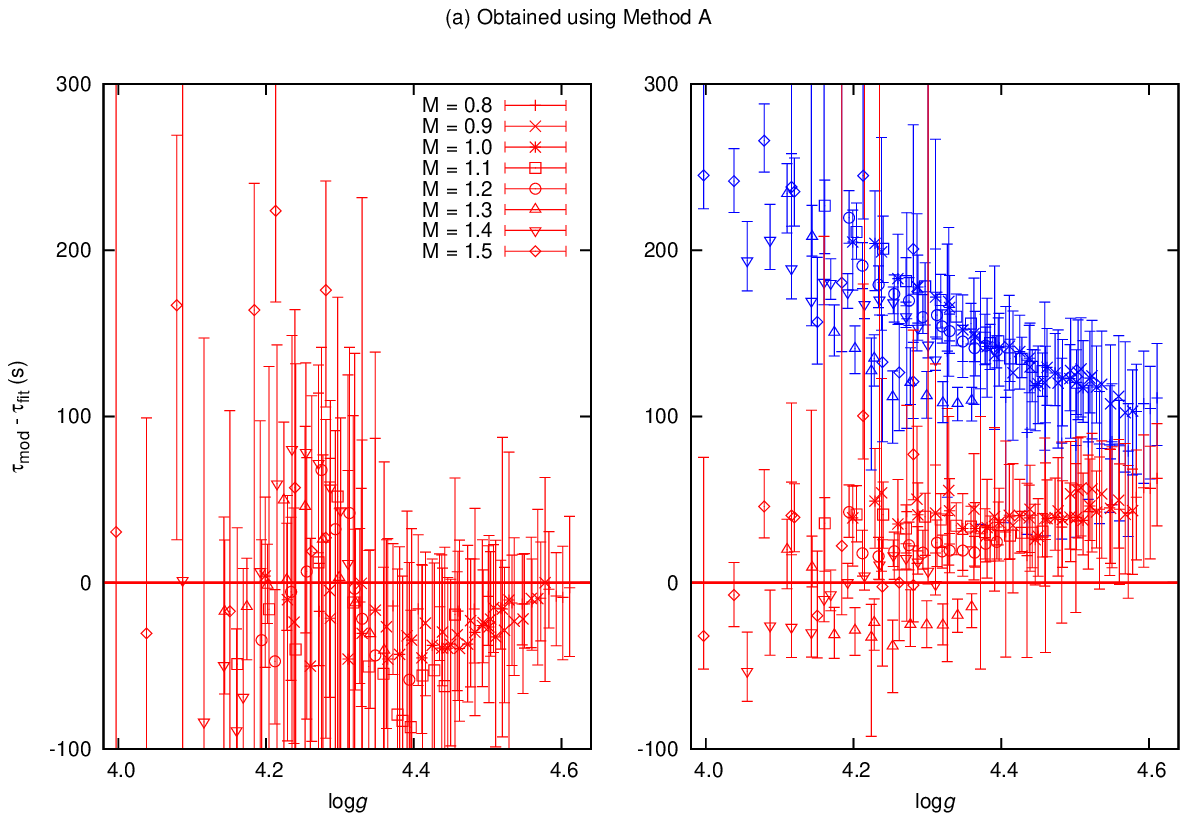}
\plotone{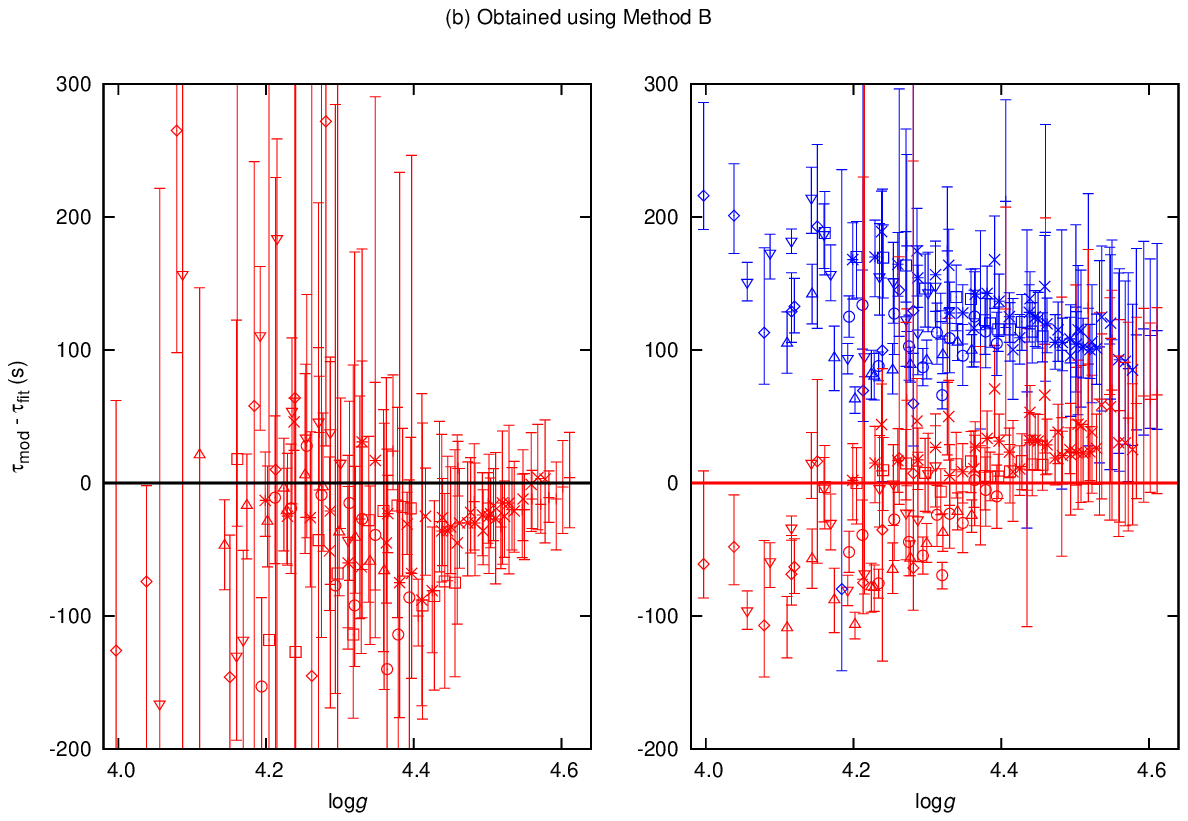}
\caption{The differences between the fitted $\tau_g$ and the acoustic depth of 
various glitches as obtained using sound speed profile. The left panels show the 
difference between the acoustic depth of the base of convection zone and 
the fitted $\tau_\mathrm{CZ}$, while the right panels show the difference 
$\tau_\mathrm{P2}-\tau_\mathrm{He}$ (red points) and 
$\tau_\mathrm{D3}-\tau_\mathrm{He}$ (blue points). Different types of points 
correspond to the masses of the stars as shown in top left panel.}
\label{fig:tau}
\end{figure}

Fig.~\ref{fig:bv} shows $N$ as a function of radius for a few 
models with masses of $1.1M_\odot$ and $1.3M_\odot$. We had no
difficulty in fitting the signal of the He ionization zones for the $1.1M_\odot$ stellar
models with age less than about 5.3 Gyr, and at these ages,
the peak in $N$ is less than about half of the lowest 
frequency used in the fits. As age increase, the peak in $N$ increases,
and we also find the quality of the fit deteriorates (as
manifest in an increased $\chi^2$ and uncertainties in the fitted parameters) even though
the fits are done using model frequencies.
A similar behavior is seen for 
models of mass less than $1.1M_\odot$, i.e., the fit becomes poor only  very late 
along the main-sequence, close to the turnoff. The scenario is different for models of
 mass greater than $1.1M_\odot$
because these have convective cores.  Most of the  models of 
mass $1.3M_\odot$ show a strong peak in $N$ just outside the convective core, 
which affects the frequencies in a manner that is not modelled by the asymptotic 
theory of stellar oscillations. We can see from Fig.~\ref{fig:bv} that all the 
models except the model with age 0.25 Gyr show a peak in $N$ that is comparable to,
or higher, than the lowest frequency used in the fitting. As a 
result, the smooth part of the frequency as a function of $n$ becomes  
complicated for low frequency modes with frequencies comparable to the maximum of 
$N$ in the core. This leads to difficulties in fitting the oscillatory signal caused by the
acoustic glitches. These low frequency modes are crucial for 
fitting the He signal because its amplitude falls off rapidly with frequency, and hence
removing these modes from the fit is not a good option. The models shown in Fig.~\ref{fig:bv}
were constructed without core overshoot. Inclusion of overshoot above
the convective core will change the models, but it is not clear if that 
will reduce $N$ just above the convective core substantially; the effect will
depend on the prescription used to include overshoot. 
In the higher mass range, 
only a few models at the beginning of the main-sequence life do not have pronounced peak in $N$, but 
these are the models where the acoustic depths of HeII ionization zone and base of convection zone 
are similar, and hence, difficult to distinguish. Thus most models in this mass range are difficult to fit.
Fits using Method B are affected more severely
because it depends on calculating the second difference of the frequencies.
It should, however, be noted that despite the difficulties in fitting the signal
and the resulting large $\chi^2$, the fitted
values of $\tau_\mathrm{He}$ and $\tau_\mathrm{CZ}$ are generally still reasonable
(see Fig.~\ref{fig:tau}),
and thus it should be possible to infer these quantities from the observed frequencies
for such stars if they are available.
\citet{mazu14} 
also had difficulty in fitting oscillatory signal to observed frequencies for some 
stars, but found reasonable values of  $\tau_\mathrm{He}$ and $\tau_\mathrm{CZ}$
for these stars.

The  panels on the left-hand-side of Fig.~\ref{fig:tau} show the difference between the acoustic depth 
of the base of the convection zone and the fitted $\tau_\mathrm{CZ}$. The small offset 
that is seen can be attributed to the choice of the stellar surface. The red and blue points in 
the right-hand-side panels show the differences $\tau_\mathrm{P2}-
\tau_\mathrm{He}$ and $\tau_\mathrm{D3}-\tau_\mathrm{He}$ respectively. 
Clearly, the peak marked  P2 corresponds more closely to the fitted $\tau_\mathrm{He}$ than the 
dip marked D3. We note from the figure, bearing in mind the offset due to the choice of 
the stellar surface, that the fitted $\tau_\mathrm{He}$ for high mass stars correspond 
very closely to a layer near the peak marked P2, and this layer moves outwards for smaller 
masses. This is expected because the fitted $\tau_\mathrm{He}$ is supposed to give 
the location of the peak in $\delta\Gamma_1$ and not in $\Gamma_1$, where $\delta\Gamma_1$ 
is the difference between the actual $\Gamma_1$ and a smooth background profile that is similar to what
is shown in Fig.~\ref{fig:gam11} as the dotted line between D2 and D3. This dotted 
line has smaller slope for more massive stars than lower mass stars with a similar helium abundance. 
Hence, the peak in $\delta\Gamma_1$ is close to the peak in $\Gamma_1$, resulting in a good 
agreement between the fitted $\tau_\mathrm{He}$ and $\tau_\mathrm{P2}$. The slope of 
the background line is larger for low mass stars thereby shifting the peak in $\delta\Gamma_1$ 
to a lower acoustic depth. We subtracted the background line for three models of mass 
$0.8M_\odot$, $1.0M_\odot$, and $1.5M_\odot$, and found the difference between the 
peak in $\Gamma_1$ and $\delta\Gamma_1$ to be 45~s, 30~s, and 15~s respectively. The 
difference between the blue and red points reflect the difference between the acoustic 
depth of P2 and D3.

\section{Conclusions}
\label{sec:disc}

In this work we have fitted oscillatory signal due to the He ionization zones
and the base of the envelope convection zone for stellar models in the mass
range of 0.8--1.5 $M_\odot$.
We first studied the Sun, 16~Cyg~A, and their representative models to investigate
 detecting a similar signal
from the HeI ionization zone and to identify the fitted acoustic depths with known features 
in the stellar models.
These stars were chosen as they represent the best case scenario for
seismic studies.
The technique was then applied to a series of stellar models.

We find  that the fitted acoustic depth of the convection zone agrees 
with that in the stellar models, while the fitted acoustic depth of the He 
ionization zone corresponds to a layer above the HeII ionization zone where 
$\Gamma_1$ is close to maximum.
Note that similar results were obtained by \citet{broo14} for models of red giants.
This contradicts the common assumption that the signal
of the acoustic glitch arises from the dip in $\Gamma_1$ caused
by the HeII ionization and one that is  used to derive
the oscillatory contribution of the glitch from the asymptotic theory of stellar oscillations 
\citep{mont05,houd07}.
The form of the oscillatory signal does not depend on 
whether it is due to a peak or a dip in $\Gamma_1$ as long as the glitch is approximately 
Gaussian in shape, and we can still use the same model to fit the oscillatory 
signal. However, we need to be careful while interpreting the results to
measure the depth of the ionization zones. We did not find any significant signal 
from the HeI ionization zone, but an attempt to fit the signal for a  solar model yielded an additional 
glitch at $\tau_g\approx170$ s, which is just above the HI ionization zone. Thus it 
appears that because of the overlap between ionization zones of HI and HeI, there is 
no peak in $\Gamma_1$ between the two and hence there is no clear oscillatory signal that
can be fitted. 
This was further verified by constructing a solar model in which the ionization
potentials of helium were increased to separate out the ionization zones. For that
model the signal due to HeI ionization zone can be fitted successfully and the $\tau_g$
obtained from fitting the signature of the glitches in the oscillation frequencies
corresponds to the peaks in the $\Gamma_1$ between the ionization zones.

The amplitude of the oscillatory signal caused by the  He ionization zones increases
with effective temperature and stellar mass.
The signal is easiest to fit for masses between 0.9--1.2 $M_\odot$.
For lower mass stars, the dip in $\Gamma_1$ in the HeII ionization zone is shallow and it is difficult to
fit the signal reliably, unless low frequency modes are included. It may
not be possible to observe these modes in the oscillation power spectrum of stars
obtained from intensity measurements.
For higher mass stars, the fit becomes unreliable because the acoustic depths
of the two glitches ($\tau_\mathrm{He},\tau_\mathrm{CZ}$) are very similar and hence difficult to
fit. Another reason for difficulty in fitting the signal for
high mass stars is the strong peak in the buoyancy frequency, $N$, just above the convective core.
This causes frequencies of modes to deviate  from the asymptotic approximation,
thus distorting the smooth part of the frequency, which needs to be
modelled appropriately to fit the signatures of the acoustic glitches. Similar difficulties 
arise even for low mass stellar models close to the end of their main-sequence life.

\acknowledgements
SB acknowledges partial support from NSF grant AST-1105930 and NASA grant
NNX13AE70G. AM acknowledges support from the NIUS programme of HBCSE (TIFR).

%\bibliographystyle{/users/kuldeep/latex/apj}
%\bibliography{/users/kuldeep/latex/references}
%

%%%%%%%%%%%%%%%%%%%%%%%%%%%%%%% Figures and Tables %%%%%%%%%%%%%%%%%%%%%%%%%%%%%%%%%%%

\clearpage
\begin{deluxetable}{cccccccccccccc}
\tabletypesize{\scriptsize}
\tablewidth{0pt}
\tablecaption{The fitted parameters for the Sun and 16 Cyg A as obtained using Method A and B.} 
\tablehead{
%&\multicolumn{6}{c}{Fit to the observed Frequencies} & &\multicolumn{6}{c}{Fit to the Model Frequencies}\\
%\cline{2-7} \cline{9-14}\\
\colhead{Method} &\colhead{$\chi^2$} &\colhead{$A_\mathrm{CZ}$} &\colhead{$\tau_\mathrm{CZ}$} 
&\colhead{$A_\mathrm{He}$} &\colhead{$\Delta_\mathrm{He}$} & \colhead{$\tau_\mathrm{He}$} \\
%& &\colhead{$\chi^2$} &\colhead{$A_\mathrm{CZ}$} &\colhead{$\tau_\mathrm{CZ}$} 
%&\colhead{$A_\mathrm{He}$} &\colhead{$\Delta_\mathrm{He}$} &\colhead{$\tau_\mathrm{He}$}\\
& &\colhead{($\mu$Hz)} &\colhead{(s)} &\colhead{($\mu$Hz)} &\colhead{(s)} &\colhead{(s)} 
%& & &\colhead{($\mu$Hz)} &\colhead{(s)} &\colhead{($\mu$Hz)} &\colhead{(s)} &\colhead{(s)}
}
\startdata
\multicolumn{7}{c}{Sun}\\[5pt]
A & 262 & $0.092 \pm 0.002$ & $2320 \pm 6$ & $0.594 \pm 0.003$ & $60.2 \pm 0.3$ & $696 \pm 1$ \\
B & 1080 & $0.080 \pm 0.002$ & $2323 \pm 4$ & $0.637 \pm 0.004$ & $61.3 \pm 0.3$& $707 \pm 1$\\
\multicolumn{7}{c}{Solar model}\\[5pt]
A & 235 & $0.099 \pm 0.002$ & $2296 \pm 5$ & $0.590 \pm 0.003$ & $61.4 \pm 0.3$ & $686 \pm 1$\\
B & 1124 & $0.085 \pm 0.002$ & $2300 \pm 4$ & $0.632 \pm 0.004$ & $61.9 \pm 0.3$ & $699 \pm 1$\\[5pt]
\multicolumn{7}{c}{16 Cyg A}\\[5pt]
A & 74.4 & $0.055 \pm 0.012$ & $3049 \pm 57$ & $0.508 \pm 0.017$ & $100.4 \pm 3.7$ & $930 \pm 13$\\ 
B & 68.9 & $0.072 \pm 0.011$ & $3079 \pm 54$ & $0.492 \pm 0.013$ & $109.0 \pm 7.0$ & $919 \pm 9$\\
\multicolumn{7}{c}{16 Cyg A model}\\[5pt]
A & 2.34 & $0.077 \pm 0.015$ & $3096 \pm 51$ & $0.506 \pm 0.018$ & $95.8 \pm 3.6$ & $883 \pm 14$\\
B & 17.8 & $0.080 \pm 0.012$ & $3098 \pm 39$ & $0.492 \pm 0.013$ & $113.9 \pm 7.8$ & $865 \pm 9$
\enddata
\label{tab:par1}
\end{deluxetable}

\clearpage
\begin{deluxetable}{ccccccccc}
\tabletypesize{\scriptsize}
\tablewidth{0pt}
\tablecaption{The fitted parameters for the solar model and the BiSON frequencies using Method A with HeI term.}
\tablehead{
\colhead{$\eta=\tau_\mathrm{I}/\tau_\mathrm{II}$} &\colhead{$\chi^2$} &\colhead{$A_\mathrm{CZ}$} &\colhead{$\tau_\mathrm{CZ}$} 
&\colhead{$A_\mathrm{I}$} &\colhead{$\Delta_\mathrm{I}$} 
&\colhead{$A_\mathrm{II}$} &\colhead{$\Delta_\mathrm{II}$} &\colhead{$\tau_\mathrm{II}$}\\
& &\colhead{($\mu$Hz)} &\colhead{(s)} &\colhead{($\mu$Hz)} &\colhead{(s)} &\colhead{($\mu$Hz)} &\colhead{(s)} &\colhead{(s)} 
}
\startdata
\multicolumn{9}{c}{Fit to the BiSON Frequencies}\\[5pt]
0.90 & 92.4 & $0.083 \pm 0.002$ & $2331 \pm 7$ & $1.023 \pm 0.098$ & $90.7 \pm 1.0$ 
& $1.417 \pm 0.075$ & $74.7 \pm 1.2$ & $631 \pm 5$\\ 
0.80 & 90.7 & $0.083 \pm 0.002$ & $2331 \pm 7$ & $0.659 \pm 0.069$ & $89.8 \pm 2.6$ 
& $1.010 \pm 0.042$ & $73.3 \pm 1.1$ & $648 \pm 5$\\ 
0.70 & 86.5 & $0.083 \pm 0.002$ & $2330 \pm 7$ & $0.540 \pm 0.076$ & $89.7 \pm 4.1$ 
& $0.827 \pm 0.038$ & $72.1 \pm 1.6$ & $660 \pm 6$\\ 
0.60 & 82.3 & $0.082 \pm 0.002$ & $2331 \pm 7$ & $0.562 \pm 0.059$ & $90.4 \pm 5.2$ 
& $0.741 \pm 0.024$ & $71.4 \pm 1.0$ & $668 \pm 5$\\ 
0.50 & 78.1 & $0.082 \pm 0.002$ & $2331 \pm 7$ & $0.704 \pm 0.060$ & $90.6 \pm 5.0$ 
& $0.693 \pm 0.017$ & $70.6 \pm 0.8$ & $675 \pm 4$\\ 
0.40 & 73.5 & $0.082 \pm 0.002$ & $2330 \pm 7$ & $1.060 \pm 0.068$ & $89.0 \pm 4.6$ 
& $0.659 \pm 0.012$ & $69.9 \pm 0.8$ & $680 \pm 4$\\ 
0.30 & 67.2 & $0.083 \pm 0.002$ & $2329 \pm 7$ & $1.964 \pm 0.124$ & $84.4 \pm 4.2$ 
& $0.632 \pm 0.007$ & $68.8 \pm 0.6$ & $686 \pm 3$\\ 
0.25 & 63.8 & $0.083 \pm 0.002$ & $2328 \pm 7$ & $2.970 \pm 0.197$ & $81.1 \pm 3.6$ 
& $0.622 \pm 0.005$ & $68.0 \pm 0.6$ & $689 \pm 2$\\ [5pt]
\multicolumn{9}{c}{Fit to the Model Frequencies}\\[5pt]
0.90 & 47.9 & $0.089 \pm 0.002$ & $2314 \pm 6$ & $1.361 \pm 0.129$ & $93.4 \pm 1.0$
& $1.669 \pm 0.100$ & $77.3 \pm 1.0$ & $605 \pm 5$\\
0.80 & 47.5 & $0.089 \pm 0.002$ & $2314 \pm 6$ & $0.892 \pm 0.092$ & $94.4 \pm 2.3$
& $1.155 \pm 0.056$ & $75.5 \pm 1.0$ & $624 \pm 5$\\
0.70 & 45.4 & $0.089 \pm 0.002$ & $2314 \pm 6$ & $0.761 \pm 0.069$ & $95.1 \pm 3.1$
& $0.933 \pm 0.032$ & $74.6 \pm 1.0$ & $636 \pm 4$\\
0.60 & 42.8 & $0.089 \pm 0.002$ & $2314 \pm 6$ & $0.783 \pm 0.071$ & $95.9 \pm 3.7$ 
& $0.816 \pm 0.024$ & $73.8 \pm 1.0$ & $646 \pm 4$\\
0.50 & 38.6 & $0.089 \pm 0.002$ & $2313 \pm 6$ & $0.938 \pm 0.076$ & $94.7 \pm 4.2$
& $0.742 \pm 0.018$ & $73.1 \pm 0.8$ & $655 \pm 4$\\
0.40 & 33.2 & $0.089 \pm 0.002$ & $2311 \pm 6$ & $1.318 \pm 0.082$ & $90.1 \pm 4.5$
& $0.685 \pm 0.013$ & $72.2 \pm 0.8$ & $664 \pm 4$\\ 
0.30 & 27.3 & $0.090 \pm 0.002$ & $2309 \pm 5$ & $2.367 \pm 0.139$ & $81.1 \pm 3.7$
& $0.639 \pm 0.008$ & $70.8 \pm 0.6$ & $674 \pm 3$\\
0.25 & 25.6 & $0.091 \pm 0.003$ & $2306 \pm 6$ & $3.573 \pm 0.403$ & $76.2 \pm 5.1$
& $0.624 \pm 0.010$ & $69.7 \pm 1.1$ & $678 \pm 4$
\enddata
\label{tab:par2}
\end{deluxetable}

\end{document}